\documentstyle{ieice1}
\FIELD{fundamentals}
\VOL{99}
\NO{99}
\YEAR{1999}
\TYPE{Preprint}
\THEME{ }
\TITLE{
Realization of a Four Parameter Family of 
Generalized One-Dimensional Contact Interactions 
by Three Nearby Delta Potentials with Renormalized Strengths
}
\authorlist{%
 \breakauthorline{3}
 \authorentry{Takaomi SHIGEHARA}{m}{labelA}
 \authorentry{Hiroshi MIZOGUCHI}{n}{labelA}
 \authorentry{Taketoshi MISHIMA}{m}{labelA}
 \authorentry{Taksu CHEON}{m}{labelB}
}
\affiliate[labelA]
{Department of Information and Computer Sciences, Saitama University, 
Shimo-Okubo, Urawa, Saitama 338-8570, Japan.} 
\affiliate[labelB]
{Laboratory of Physics, Kochi University of Technology, 
Tosa Yamada, Kochi 782-8502, Japan.}

\received{December 3, 1998.}

\begin{document}
\maketitle
\setcounter{page}{1}

\begin{SUMMARY}
We propose a new method to construct a four parameter family of 
quantum-mechanical point interactions in one dimension, which is 
known as all possible self-adjoint extensions of the symmetric 
operator $T=-\Delta \lceil C^{\infty}_{0}({\bf R} \backslash\{0\})$. 
It is achieved in the small distance limit of equally spaced 
three neighboring Dirac's $\delta$ potentials. 
The strength for each $\delta$ is appropriately renormalized 
according to the distance and it diverges, in general, in the 
small distance limit. 
The validity of our method is ensured by numerical calculations. 
In general cases except for usual $\delta$, 
the wave function discontinuity appears around the interaction 
and one can observe such a tendency 
even at a finite distance level. 
\end{SUMMARY}


\section{Introduction} 

Point interaction is a simple but useful object for examining 
the effect of small impurities on low-energy dynamics in quantum mechanics. 
It serves as a minimal perturbation which gives light 
on the influence of more complicated perturbations 
on dynamical nature of the system. 
It has been first observed \cite{SE90,AS91} that 
wave chaos (chaotic nature in quantum spectrum) is induced 
by point perturbation in two-dimensional integrable billiards 
and a general condition for its emergence has been clarified 
in \cite{SH94,SC97,SM97}. 
In \cite{SM98b}, the dependence of the ``degree'' of chaos 
on the number of point impurities has been examined in details. 

Historically, the first influential paper on point interactions 
was given by Kronig and Penney \cite{KP31}. 
The Kronig-Penney model (potential consisting of a 
periodic array of $\delta$ functions)  
has been widely considered as a 
standard reference model in solid state physics for more than six decades. 
In two and three dimensions, the point interaction is nothing but 
an extended object of the usual $\delta$ potential.  
However the phenomena are much richer in one dimension. 
This is due to the fact that whereas the symmetric operator 
$T=-\Delta \lceil C^{\infty}_{0}({\bf R}^d \backslash\{0\})$ has 
deficiency indices $(1,1)$ in spatial dimension $d=2,3$, 
it has deficiency indices $(2,2)$ in dimension $d=1$ \cite{AG88}. 
As a result, the operator $T$ has in one dimension a four-parameter family of 
self-adjoint extensions \cite{RS75}. 
The general connection condition is characterized by 
\begin{eqnarray}
\label{eq1-1}
\left( \! \!
\begin{array}{c}
\varphi'(+0) \\
\varphi (+0) 
\end{array}
\! \! \right)
=
{\cal V}
\left( \! \!
\begin{array}{c}
\varphi'(-0) \\
\varphi (-0) 
\end{array}
\! \! \right),   
\end{eqnarray}
where ${\cal V}$ takes a form  
\begin{eqnarray}
\label{eq1-2}
{\cal V}=e^{i\theta}{\cal U}
\end{eqnarray}
with $\theta \in {\bf R}$ and 
\begin{eqnarray}
\label{eq1-3}
{\cal U} = \left(
\begin{array}{cc}
\alpha & \beta \\
\gamma & \delta 
\end{array}
\right) \in SL(2,{\bf R}),   
\end{eqnarray}
namely, $\alpha\delta-\beta\gamma=1$ \cite{GK85}. 
Here we assume that the point interaction is placed at 
the origin on $x$-axis. 
The condition (\ref{eq1-1}) covers the usual $\delta$ potential, 
the connection matrix of which is given by 
\begin{eqnarray}
\label{eq1-4}
{\cal V} = {\cal V}_{\delta}(v) \equiv 
\left(
\begin{array}{cc}
1 & v \\
0 & 1 
\end{array}
\right).   
\end{eqnarray}
Here $v$ is the potential strength. 
Another interesting example is a so-called 
(historically ill-called) $\delta'$ potential, 
which induces such a boundary condition that 
the wave function has continuous first derivative 
on the right and left, but it has a jump 
proportional to the first derivative \cite{GHM80}; 
\begin{eqnarray}
\label{eq1-5}
{\cal V} = {\cal V}_{\varepsilon}(u) \equiv 
\left(
\begin{array}{cc}
1 & 0  \\
u & 1 
\end{array}
\right),  
\end{eqnarray}
where $u$ is regarded as the strength of the interaction. 
It has been shown \cite{S86b} that 
the connection condition (\ref{eq1-5}) has 
little to do with the derivative of $\delta$ potential. 
In this paper, 
we denote the contact interaction with the transfer matrix 
(\ref{eq1-5}) by $\varepsilon(x)$.  

Except for $\delta$, the connection condition 
(\ref{eq1-1}) appears at first sight to be unnatural for 
quantum mechanics, and some efforts have been made 
in order to seek the physical Hamiltonians which embody 
the condition (\ref{eq1-1}). 
\v{S}eba has observed \cite{SE86a} that a certain two-parameter 
family is obtained by non-local pseudopotentials 
with $\varepsilon$ potential. 
Chernoff and Hughes have shown \cite{CH93} that a three-parameter family 
disjoint from \v{S}eba's class corresponds to local pseudopotentials 
involving the so-called $\delta^2$. 
Carreau has succeeded in realizing a general class  
in the small-size limit of non-symmetric operators \cite{C93}. 
A local realization has been given by 
Rom\'{a}n and Tarrach by using a potential consisting of 
a step-like barrier sandwiched in two $\delta$ potentials \cite{RT96}.  
Their model requires, however, a subtle limiting procedure 
and indeed the height of the middle barrier as well as the strengths of 
the two side $\delta$'s are necessary to be renormalized with a doubly 
logarithmic dependence on the distance between the two $\delta$'s.  
In spite of considerable works, 
existing approximations do not seem to have much relevance to the 
experimentally realizable systems. 
Our main purpose is to rectify this situation. 
In \cite{CS98a,SM98a}, we have constructed the 
$\varepsilon$ potential in the small distance limit of 
three nearby $\delta$ potentials and realized 
in terms of usual $\delta$ and $\varepsilon$ potentials 
a three-parameter family of self-adjoint extensions 
under the assumption of time reversal symmetry. 
In this paper, we generalize the previous formalism and 
give in a direct manner within the three $\delta$'s model  
a complete solution for the four-parameter extensions 
covering the cases without the time reversal symmetry. 

In Sect.2, we start by solving the one-dimensional 
Schr\"{o}dinger equation with a constant vector potential 
within the framework of the transfer matrix formalism. 
Considering the potential which consists of equally spaced three $\delta$'s 
with a constant vector potential between the two side $\delta$'s, 
we show that all possible extensions are attained in the short range  
limit of the assumed potential 
if the strengths of the $\delta$'s as well as that of vector potential 
are renormalized according to the distance in a simple manner. 
In Sect.3, We give numerical examples which justify the method 
in the previous section. 
The current work is summarized in Sect.4. 

\section{Approximation of Four-Parameter Family of One-Dimensional 
Point Interactions by Three Nearby Delta Interactions} 

We first consider the Hamiltonian 
with a homogeneous (everywhere constant) vector potential 
in one dimension. It is given by a usual minimal coupling to 
the magnetic field;  
\begin{eqnarray}
\label{eq2-1}
H = (p-A)^2, 
\end{eqnarray}
where $p=-id/dx$ is the momentum operator and 
constant $A$ is the strength of the vector potential. 
The Schr\"{o}dinger equation for the Hamiltonian (\ref{eq2-1}) reads 
\begin{eqnarray}
\label{eq2-2}
-\frac{d^2 \varphi(x)}{dx^2} + 2iA \frac{d\varphi(x)}{dx} + A^2 \varphi(x) = 
k^2 \varphi(x). 
\end{eqnarray}
Here $k$ is the wave number of the particle. 
For later convenience, we rewrite the equation (\ref{eq2-2}) 
within the transfer matrix formalism. 
For this purpose, let us introduce a vector notation for 
the wave function and its space derivative; 
\begin{eqnarray}
\label{eq2-3}
 {\bf \Psi}(x)= \left( \! 
\begin{array}{c}
\varphi'(x) \\
\varphi(x) \\
\end{array}
\! \right).   
\end{eqnarray}
In this notation, the equation (\ref{eq2-2}) is rewritten by 
the first-order coupled equation  
\begin{eqnarray}
\label{eq2-4}
{\bf \Psi}'(x) = {\cal H}(A,k)
{\bf \Psi}(x)
\end{eqnarray}
with
\begin{eqnarray}
\label{eq2-5}
{\cal H}(A,k) = 
\left(
\begin{array}{cc}
2iA & -k^2 + A^2  \\
1   & 0
\end{array}
\right). 
\end{eqnarray}
The solution of Eq.(\ref{eq2-4}) is given by 
\begin{eqnarray}
\label{eq2-6}
{\bf \Psi}(x) = {\cal G}(A,k;x-x_0) {\bf \Psi}(x_0), 
\end{eqnarray}
where ${\cal G}(A,k;x)$ is the exponential function of ${\cal H}(A,k)x$; 
\begin{eqnarray}
\label{eq2-7}
\hspace*{-3ex}
{\cal G}(A,k;x) & \equiv & e^{{\cal H}(A,k)x} \nonumber \\ 
\hspace*{-3ex}
& = & e^{iAx} \left\{
\cos ( kx ) 
\left(
\begin{array}{cc}
 1 & 0 \\
 0 & 1
\end{array}
\right) 
\right. \nonumber \\ 
\hspace*{-3ex}
&  & +  \left. 
\frac{\sin ( kx )}{k}
\left(
\begin{array}{cc}
iA & -k^2 + A^2\\
 1 & -iA
\end{array}
\right) 
\right\}. 
\end{eqnarray}
Since 
\begin{eqnarray}
\label{eq2-8}
Tr {\cal H}(A,k)=2iA, 
\end{eqnarray}
the matrix ${\cal G}(A,k;x)$ satisfies 
\begin{eqnarray}
\label{eq2-9}
\det {\cal G}(A,k;x) = e^{2iAx}, 
\end{eqnarray}
indicating that the matrix on the RHS of Eq.(\ref{eq2-7}) 
has determinant one. 

\setlength{\unitlength}{1mm}
\begin{figure}[t]
\begin{center}
\begin{picture}(67,60)
\thicklines 
\put(5,10){\vector(1,0){57}} 
\put(8,5){\vector(0,1){47}}
\put(64,9){\Large $x$}
\put(0,55){\Large $V_a(x)$}
\put(35,10){\line(0,1){25}}
\put(34,6){\large $0$}
\put(30,38){\large $v_0(a)$}
\put(25,10){\line(0,-1){3.5}}
\put(21,12){\large $-a$}
\put(20,1){\large $v_-(a)$}
\put(45,10){\line(0,-1){5}}
\put(44,12){\large $a$}
\put(40,1){\large $v_+(a)$}
\multiput(25,15)(0,2){18}{\line(0,1){1}}
\multiput(45,15)(0,2){18}{\line(0,1){1}}
\put(9.5,45){$A_a(x)=0$}
\put(27.5,45){$A_a(x)=A$}
\put(47,45){$A_a(x)=0$}
\end{picture}
\end{center}
\hspace*{1ex}
\begin{center}
\parbox{8cm}{
{\small \setlength{\baselineskip}{10pt}%
{\bf Fig.1} Approximation of a general class of 
point interactions by three neighboring $\delta$ 
potentials. The vector potential $A_a(x)$ has a step-like 
form: a constant $A$ between two side $\delta$'s 
and otherwise zero. 
The strengths of the three $\delta$'s depend on 
the distance $a$ in general. It will be shown 
that if 
$v_+(a) = -1/a + (\alpha+1)/\gamma$, 
$v_0(a) = \gamma/a^2$ and 
$v_-(a) = -1/a + (\delta+1)/\gamma$ 
together with $\theta = 2Aa$ constant,   
one attains in the small $a$ limit 
the boundary condition (\ref{eq1-1}) for $\gamma\neq 0$. 
For $\gamma=0$, the condition is realized if  
$v_+(a) = (\alpha-1)/(2a)$, 
$v_0 = 4\beta/(\alpha+\delta+2)$ and 
$v_-(a) = (\delta-1)/(2a)$,   
respectively. 

}}
\end{center}
\end{figure}

In order to realize the general condition (\ref{eq1-1}) 
in the small-size limit 
of a realistic finite-range potential, 
we consider three neighboring usual $\delta$'s 
which are located with equal distance $a$ \cite{CS98a}. 
In order to take into account the time reversal symmetry breaking, 
we add a constant vector potential between the two side $\delta$'s; 
\begin{eqnarray}
\label{eq2-10}
A_a(x) = A \theta(a-x) \theta(x+a), 
\end{eqnarray}
where $\theta$ is a usual step function. 
The Hamiltonian is hence given by 
\begin{eqnarray}
\label{eq2-11}
H_a & = & \left( p-A_a(x) \right)^2 + v_- \delta(x+a) 
+ v_0 \delta(x) \nonumber \\ 
&  & + v_+ \delta(x-a),  
\end{eqnarray}
where the strengths of the three delta's at $x=\pm a$ and $0$ are set to 
$v_{\pm}$ and $v_{0}$, respectively.  
The assumed potential $V_a(x)$ is shown in Fig.1. 
Inserting Eq.(\ref{eq2-10}) into Eq.(\ref{eq2-11}), one obtains 
\begin{eqnarray}
\label{eq2-12}
H_a & = & -\frac{d^2}{dx^2} + 2iA_a(x) \frac{d}{dx} + A_a(x)^2  
\nonumber \\
& & + (v_- + iA) \delta(x+a) +  v_0 \delta(x) \nonumber \\ 
& & + (v_+ - iA) \delta(x-a).    
\end{eqnarray}
The appearance of the complex $\delta$'s at $x=\pm a$ are 
inevitable for ensuring the Hamiltonian to be symmetric and 
they indeed describe a sudden change of the vector potential 
at the location of the two side $\delta$'s.
With the Hamiltonian (\ref{eq2-12}),  
the connection of ${\bf \Psi}$ 
between $x=-a-0$ and $x=a+0$ is given by 
\begin{eqnarray}
\label{eq2-13}
{\bf \Psi}(a+0) = {\cal V}_{a}(A,k){\bf \Psi}(-a-0), 
\end{eqnarray}
where
\begin{eqnarray}
\label{eq2-14}
{\cal V}_{a}(A,k) & = &
{\cal V}_{\delta}(v_+ - iA)
{\cal G}(A,k;a) 
{\cal V}_{\delta}(v_0) \nonumber \\
& & \times {\cal G}(A,k;a) 
{\cal V}_{\delta}(v_- + iA). 
\end{eqnarray}
Inserting Eqs.(\ref{eq1-4}) and (\ref{eq2-7}) into Eq.(\ref{eq2-14}), 
we obtain 
\begin{eqnarray}
\label{eq2-15}
{\cal V}_{a}(A,k) = e^{2iAa} {\cal U}_{a}(k), 
\end{eqnarray}
where the $(2,1)$ component of ${\cal U}_{a}(k)$ is given by 
\begin{eqnarray}
\label{eq2-16}
\left[{\cal U}_{a}(k)\right]_{21} = \frac{\sin ( 2ka )}{k} + 
\frac{ \sin^2 ( ka )}{k^2} v_0 
\end{eqnarray}
and the other elements are written as  
\begin{eqnarray}
\label{eq2-17}
\hspace*{-5ex}
\left[{\cal U}_{a}(k)\right]_{11} & = & 
\cos ( 2ka ) + \frac{\sin ( 2ka )}{2k} v_0 
+ \left[{\cal U}_{a}(k)\right]_{21} v_+, \\
\label{eq2-18}
\hspace*{-5ex}
\left[{\cal U}_{a}(k)\right]_{22} & = & 
\cos ( 2ka ) + \frac{\sin ( 2ka )}{2k} v_0 
+ \left[{\cal U}_{a}(k)\right]_{21} v_-, \\
\label{eq2-19}
\hspace*{-5ex}
\left[{\cal U}_{a}(k)\right]_{12} & = & 
\cos^2 ( ka ) \cdot (v_+ + v_0 + v_-) \nonumber \\
\hspace*{-5ex}
& & - \sin^2 ( ka ) \cdot (v_+ + v_-) \nonumber \\
\hspace*{-5ex}
& & + \frac{\sin ( 2ka )}{2k} 
\left\{ -2k^2 + v_0 ( v_+ + v_-) \right\} \nonumber \\
& & + \left[{\cal U}_{a}(k)\right]_{21} v_+ v_- 
\end{eqnarray}
in terms of $\left[{\cal U}_{a}(A,k)\right]_{21}$, respectively. 
Note that the effect of the magnetic field is absorbed in 
the phase factor in front of the matrix 
${\cal U}_{a}(k)$ in Eq.(\ref{eq2-15}). 
Since 
\begin{eqnarray}
\label{eq2-20}
\det {\cal V}_{a}(A,k) = e^{4iAa} 
\end{eqnarray}
from Eqs.(\ref{eq2-9}) and (\ref{eq2-14}), 
the real matrix ${\cal U}_{a}(k)$ satisfies 
\begin{eqnarray}
\label{eq2-21}
\det {\cal U}_{a}(k) = 1, 
\end{eqnarray}
namely, ${\cal U}_{a}(k) \in SL(2,{\bf R})$. 

The phase factor of the matrix (\ref{eq2-15}) shows that  
in order to realize the general connection condition (\ref{eq1-1}), 
one has to take a small-size limit under the condition 
\begin{eqnarray}
\label{eq2-22}
2Aa = \theta 
\end{eqnarray}
with a fixed $\theta \in {\bf R}$. 
Accordingly, the strength of the vector potential increases as 
$a \longrightarrow +0$;    
\begin{eqnarray}
\label{eq2-23}
A(a) = \frac{\theta}{2a}. 
\end{eqnarray}
As shown just below, 
a singular behavior for small $a$ 
is demanded for the strengths of the three $\delta$'s potentials  
in order to attain the condition (\ref{eq1-1}).  
This enforces a careful analysis of each matrix element of ${\cal U}_{a}(k)$ 
in the small distance limit. 
Indeed, it requires the expansions of triangular functions appeared in 
Eqs.(\ref{eq2-16}) -- (\ref{eq2-19}) up to the orders shown below; 
\begin{eqnarray}
\label{eq2-24}
\cos( 2ka ) & = & 1 + O(a^2), \\
\label{eq2-25}
\sin ( 2ka ) & = & 
2ka - \frac{4}{3}k^3 a^3 + O(a^5), \\
\label{eq2-26}
\cos^2 ( ka ) & = & 1 - k^2 a^2 + O(a^4), \\
\label{eq2-27}
\sin^2 ( ka ) & =  & k^2 a^2 - \frac{1}{3}k^4 a^4 + O(a^6). 
\end{eqnarray} 
Inserting Eqs.(\ref{eq2-25}) and (\ref{eq2-27}) 
into Eq.(\ref{eq2-16}), one observes 
\begin{eqnarray}
\label{eq2-28}
\left[{\cal U}_{a}(k)\right]_{21} =  2a + v_0 a^2  + O(a^3).       
\end{eqnarray}
This indicates that the strength $v_0$ of the middle $\delta$ 
is required to diverge with the order of $1/a^2$ in the small distance limit 
in order to make $\left[{\cal U}_{a}(k)\right]_{21}$ have a non-zero limit. 
Indeed, if $v_0$ is changed according to the 
distance $a$ as 
\begin{eqnarray}
\label{eq2-29}
v_0(a) = \frac{\gamma}{a^2} 
\end{eqnarray}
with an arbitrary real constant $\gamma \neq 0 $, 
one obtains 
\begin{eqnarray}
\label{eq2-30}
\left[{\cal U}_{a}(k)\right]_{21} & = & \gamma + 2a 
- \frac{\gamma}{3}k^2 a^2 + O(a^3),    
\end{eqnarray}
leading to 
\begin{eqnarray}
\label{eq2-31}
\lim_{a \longrightarrow +0}
\left[{\cal U}_{a}(k)\right]_{21} = \gamma.   
\end{eqnarray}
Using Eq.(\ref{eq2-30}), one has the estimates 
\begin{eqnarray}
\label{eq2-32}
\left[{\cal U}_{a}(k)\right]_{11} & = & 
\frac{\gamma}{a} + 1 + \left( \gamma + 2a + O(a^2) \right) v_+ 
\nonumber \\ 
& & + O(a), \\
\label{eq2-33}
\left[{\cal U}_{a}(k)\right]_{22} & = & 
\frac{\gamma}{a} + 1 + \left( \gamma + 2a + O(a^2) \right) v_- 
\nonumber \\
& & + O(a) 
\end{eqnarray}
for small $a$. 
The RHS of Eqs.(\ref{eq2-32}) and (\ref{eq2-33}) 
diverge in the small $a$ limit if 
one keeps $v_+$ and $v_-$ constant. 
The strengths $v_+$, $v_-$ of the side $\delta$'s are 
demanded to diverge with the order of $1/a$ 
to make $\left[{\cal U}_{a}(k)\right]_{11}$ and 
$\left[{\cal U}_{a}(k)\right]_{22}$ 
converge in the small $a$ limit;  
If 
\begin{eqnarray}
\label{eq2-34}
v_+(a) & = & -\frac{1}{a} + \frac{\alpha+1}{\gamma}, \\
\label{eq2-35}
v_-(a) & = & -\frac{1}{a} + \frac{\delta+1}{\gamma} 
\end{eqnarray}
with real constants $\alpha$ and $\delta$, 
the diagonal elements converge into 
\begin{eqnarray}
\label{eq2-36}
\lim_{a \longrightarrow +0}
\left[{\cal U}_{a}(k)\right]_{11} & = & \alpha, \\
\label{eq2-37}
\lim_{a \longrightarrow +0}
\left[{\cal U}_{a}(k)\right]_{22} & = & \delta, 
\end{eqnarray}
respectively. 
Note that the divergent term in $v_+$ and $v_-$ is 
dependent only on the distance $a$ between neighboring $\delta$'s 
and the regular part determines the boundary condition 
around the generalized point interaction. 
Since ${\cal U}_{a}(k) \in SL(2,{\bf R})$, one obtains 
under the constraints (\ref{eq2-29}), (\ref{eq2-34}) 
and (\ref{eq2-35}) 
\begin{eqnarray}
\label{eq2-38}
\lim_{a \longrightarrow +0}
\left[{\cal U}_{a}(k)\right]_{12} = 
\frac{\alpha\delta-1}{\gamma} = \beta,  
\end{eqnarray}
which can be shown in a direct manner by noticing 
Eqs.(\ref{eq2-25})--(\ref{eq2-27}) and (\ref{eq2-30}). 
Since $v_0=O(1/a^2)$, $v_+=O(1/a)$ and $v_-=O(1/a)$ 
respectively, a close examination of the higher order terms 
in each equation is required to observe 
the cancellation among the terms of $O(1/a^2)$, $O(1/a)$ and $O(a^0$).  

For the $\varepsilon$ potential of strength $u$,   
we have 
\begin{eqnarray}
\label{eq2-39}
v_0(a) & = & \frac{u}{a^2}, \\
\label{eq2-40}
v_+(a) & = & v_-(a) = -\frac{1}{a} + \frac{2}{u}, 
\end{eqnarray}
which is exactly what we have shown in \cite{SM98a}. 

Up to now, we have excluded the case of $\gamma = 0$. 
For this case, 
we start by assuming $v_0$ to be constant (independent of 
the distance $a$), which is appropriately determined later. 
With this assumption, we see from Eq.(\ref{eq2-16}) 
\begin{eqnarray}
\label{eq2-41}
\gamma = \lim_{a \longrightarrow +0}
\left[{\cal U}_{a}(k)\right]_{21} = 0. 
\end{eqnarray}
Moreover, Eqs.(\ref{eq2-17}) and (\ref{eq2-18}) are reduced to 
\begin{eqnarray}
\label{eq2-42}
\left[{\cal U}_{a}(k)\right]_{11} & = & 
1 + \left( 2a + O(a^2) \right) v_+ + O(a), \\
\label{eq2-43}
\left[{\cal U}_{a}(k)\right]_{22} & = & 
1 + \left( 2a + O(a^2) \right) v_- + O(a),  
\end{eqnarray}
respectively. 
Eqs.(\ref{eq2-42}) and (\ref{eq2-43}) indicate that if 
\begin{eqnarray}
\label{eq2-44}
v_+ (a) = \frac{\alpha-1}{2a}, \\
\label{eq2-45}
v_- (a) = \frac{\delta-1}{2a} 
\end{eqnarray}
for small $a$, we attain 
\begin{eqnarray}
\label{eq2-46}
\lim_{a \longrightarrow +0}
\left[{\cal U}_{a}(k)\right]_{11} & = & \alpha, \\
\label{eq2-47}
\lim_{a \longrightarrow +0}
\left[{\cal U}_{a}(k)\right]_{22} & = & \delta = 1/\alpha.  
\end{eqnarray}
Here $\alpha\delta=1$ is imposed since $\det {\cal U}_a(k)=1$ 
and $\gamma=0$. 
It is easy to see from Eqs.(\ref{eq2-44}) and (\ref{eq2-45}) 
that the condition $\alpha\delta=1$ is 
equivalent to 
\begin{eqnarray}
\label{eq2-48}
v_+ + v_- + 2 v_+ v_- a = 0.  
\end{eqnarray}
Using 
Eqs.(\ref{eq2-28}), (\ref{eq2-44}), (\ref{eq2-45}) and (\ref{eq2-48}), 
we have 
\begin{eqnarray}
\label{eq2-49}
\hspace*{-3ex}
v_+ + v_- + 
\left[{\cal U}_{a}(k)\right]_{21} v_+ v_- & = & 
\frac{v_0(\alpha-1)(\delta-1)}{4} \nonumber \\  
\hspace*{-3ex}
& & + O(a). 
\end{eqnarray}
We thus obtain the estimate of Eq.(\ref{eq2-19}) 
for small $a$;  
\begin{eqnarray}
\label{eq2-50}
\left[{\cal U}_{a}(k)\right]_{12} & = & 
v_+ + v_- + v_0  + v_0(v_+ + v_-) a \nonumber \\
& & + \left[{\cal U}_{a}(k)\right]_{21} v_+ v_- + O(a) \nonumber \\
& = & \frac{\alpha+\delta+2}{4} v_0 +O(a).  
\end{eqnarray}
Eq.(\ref{eq2-50}) shows that if the constant $v_0$ is taken as 
\begin{eqnarray}
\label{eq2-51}
v_0 = \frac{4\beta}{\alpha+\delta+2} 
\end{eqnarray}
with an arbitrary real $\beta$, one obtains the limit 
\begin{eqnarray}
\label{eq2-52}
\lim_{a \longrightarrow +0}
\left[{\cal U}_{a}(k)\right]_{12} & = & \beta. 
\end{eqnarray}
This completes the case of $\gamma=0$. 
In a particular case of $\alpha=\delta=1$ and $\beta=v$,     
Eqs.(\ref{eq2-44}), (\ref{eq2-45}) and (\ref{eq2-51}) are reduced to 
$v_+ =  v_- = 0$ and $v_0 =  v$, namely   
a single $\delta$ potential of strength $v$ as expected. 
[Strictly speaking, Eq.(\ref{eq2-51}) excludes the case of 
$\alpha=\delta=-1$; ${\cal U}=-{\cal V}_{\delta}(-\beta)$. 
However, this is easily realized by using a single $\delta$ 
($v_0=-\beta$, $v_+=v_-=0$) together with 
the phase replacement from $\theta$ to $\theta+\pi$.] 

Our findings in this section are summarized as follows.  
The Hamiltonian $H_a$ in Eq.(\ref{eq2-12}) produces 
all possible connection conditions in Eq.(\ref{eq1-1}) 
in the small $a$ limit  
together with suitably renormalized strengths; 
In case of $\gamma\neq 0$, 
Eqs.(\ref{eq2-34}), (\ref{eq2-35}), 
(\ref{eq2-29}) and  (\ref{eq2-23}) 
for $v_+$, $v_-$, $v_0$ and $A$ respectively,  
whereas in the case $\gamma=0$,   
Eqs.(\ref{eq2-44}), (\ref{eq2-45}) and (\ref{eq2-51}) 
for $v_+$, $v_-$ and $v_0$ together with Eq.(\ref{eq2-23}) for $A$. 

\section{Numerical Examples}

\begin{figure}[t]
\begin{center}
  \input{wf2_10.tex}
  \input{wf2_13.tex}
\end{center}
\hspace*{1ex}
\begin{center}
\parbox{8cm}{
{\small \setlength{\baselineskip}{10pt}%
{\bf Fig.2} 
The solid line shows the wave function of the tenth (upper) and 
the thirteenth (lower) eigenstate  
for the generalized point interaction with 
$\alpha=3$, $\beta=-2$, $\gamma=-7$ and $\delta=5$. 
The wave number is $k_{10}=0.894964$ and $k_{13}=1.130869$, respectively. 
For comparison, 
the approximated wave function with three $\delta$'s potential 
with the distance $a=0.2$ is exhibited by the broken line. 
In this case, the wave number is 
$k_{10}=0.905264$ and $k_{13}=1.142775$, respectively. 

}}
\end{center}
\end{figure}

In order to show the validity of the three $\delta$'s 
approximation, we perform numerical tests in this section. 
For simplicity, we consider the bounded region $[X_1,X_2]$ ($X_1 < 0 < X_2$) 
together with the Dirichlet boundary condition 
$\varphi(X_1)=\varphi(X_2)=0$. 
The eigenvalues $k_n$ for the generalized point interaction 
are determined by 
\begin{eqnarray}
\label{eq3-1}
\left[ {\cal G}(0,k_n;X_2){\cal V}{\cal G}(0,k_n;-X_1) \right]_{21}=0  
\end{eqnarray}
with Eqs.(\ref{eq1-2}) and (\ref{eq2-7}). 
The associated (not necessarily normalized) eigenfunction is 
calculated by 
\begin{eqnarray}
\label{eq3-2}
\hspace*{-5ex}
\varphi_n(x)= \left\{ 
\begin{array}{ll}
\left[ {\cal G}(0,k_n;x-X_1) \right]_{21}, & \hspace*{-1ex} x<0, \\
\left[ {\cal G}(0,k_n;x){\cal V}{\cal G}(0,k_n;-X_1) \right]_{21}, & 
\hspace*{-1ex} x>0.  
\end{array}
\right. 
\end{eqnarray}
The eigenvalues for the three $\delta$'s approximation  
are determined by 
\begin{eqnarray}
\label{eq3-3}
\hspace*{-5ex}
\left[ 
{\cal G}(0,k_n;X_2-a)
{\cal V}_{a}(A,k_n) 
{\cal G}(0,k_n;-a-X_1) \right]_{21}=0  
\end{eqnarray}
with Eq.(\ref{eq2-14}), where $A=\theta/(2a)$ according to Eq.(\ref{eq2-23}). 
In Eq.(\ref{eq2-14}), the strengths of three $\delta$'s are 
determined by Eqs.(\ref{eq2-29}), (\ref{eq2-34}), (\ref{eq2-35}) 
for $\gamma\neq 0$ and 
Eqs.(\ref{eq2-44}), (\ref{eq2-45}), (\ref{eq2-51}) 
for $\gamma=0$ respectively. 
The associated wave function is calculated in a similar manner 
to in Eq.(\ref{eq3-2}). 
In the following, 
we set the vector potential $A=0$ ($\theta=0$) since its effect is trivial. 
In the numerical calculations, we take the endpoints 
$X_2=-X_1=15.0$ and the distance $a=0.2$. 

Figure 2 shows the eigenfunction of two low-energy 
(tenth in upper and thirteenth in lower) eigenstates 
for the case of $\alpha=3$, $\beta=-2$, $\gamma=-7$ and $\delta=5$. 
The solid line is a calculation for the point interaction, 
whereas the broken line is the wave function approximated 
by three nearby $\delta$'s. 
Figure 3 shows a typical example for the case of $\gamma=0$. 
We take $\alpha=5$, $\beta=3$, $\delta=0.2$. 
In all cases, 
one can observe the wave function discontinuity around the point 
interaction,  
which is successfully reproduced in the corresponding eigenfunction 
approximated by three $\delta$'s potential with appropriately 
renormalized strengths.  
However, it is worthy to mention that 
the approximated $\varphi_n$ always has $n-1$ nodes on the intervals 
as long as the distance $a$ is kept finite, whereas this is not 
necessarily the case in the limit of $a\longrightarrow +0$, 
namely for the case of the generalized point interaction. 
Indeed, the lower part in Fig.2 shows that $\varphi_{13}$ 
for the exactly point case loses two nodes in the vicinity of the 
interaction and as a result it has only ten nodes on the interval 
$(X_1,X_2)$. On the other hand, the approximated (continuous) 
wave function behaves around the three $\delta$'s in  
a somewhat complicated manner to ensure the relation between 
the number of nodes and the quantum number $n$ mentioned above. 
Clearly, the convergence of the wave function in the small $a$ limit 
is not uniform on the interval including the origin in general. 

\begin{figure}[bt]
\begin{center}
  \input{wf3_7.tex}
\end{center}
\hspace*{1ex}
\begin{center}
\parbox{8cm}{
{\small \setlength{\baselineskip}{10pt}%
{\bf Fig.3} 
The solid line shows the wave function 
of the seventh eigenstate ($k_7=0.775671$ ) 
for the generalized point interaction with 
$\alpha=5$, $\beta=3$, $\gamma=0$ and $\delta=0.2$. 
For comparison, 
the approximated wave function ($k_7=0.775312$ ) 
by three $\delta$'s potential 
with the distance $a=0.2$ 
is exhibited by the broken line. 

}}
\end{center}
\end{figure}

\section{Conclusion}

We have constructed 
one-dimensional generalized point interactions characterized 
by four parameters 
in the small distance limit of equally spaced three $\delta$'s potentials. 
A constant vector field added on the interval between the two side
$\delta$'s  
has no effect except for changing the phase of wave function. 
The remaining part, the transfer matrix ${\cal U}_a(k)$ can be made converge 
into an arbitrarily fixed special linear matrix by adjusting the strengths 
of the three $\delta$'s according to the distance in an appropriate manner. 
Though a divergent term is inevitable for each strength 
in the small distance limit,  
it takes a remarkably simple and experimentally realizable form. 
Numerical examples support our model and show a satisfactory coincidence 
with the corresponding zero-range case even at a finite distance level.

\profile{Takaomi Shigehara}{%
received the B.S, M.S and Ph.D. degrees in Physics 
from the University of Tokyo in 1983, 1985 and 1988, 
respectively. 
He is currently an Assistant Professor in the Department of 
Information and Computer Sciences at Saitama University. 
His research interests are quantum chaos, high-performance computing, 
and numerical analysis. 
}

\profile{Hiroshi Mizoguchi}{%
received the B.E. degree in Mathematical Engineering \linebreak in 1980, 
and the M.E. and Ph.D. degrees in Information Engineering 
in 1982 and 1985, respectively, from the University of Tokyo.  
He is currently an Associate Professor in the Department of 
Information and Computer Sciences at Saitama University. 
His research interests are vision processor system, robotics, 
quantum chaos, and parallel computation.  
} 

\profile{Taketoshi Mishima}{%
received the B.E., M.E. and Ph.D. degrees in 
Electrical Engineering 
from Meiji University, Japan, in 1968, 1970 and 1973, 
respectively. 
He is currently a Professor in the Department of 
Information and Computer Sciences at Saitama University. 
His research interests are foundation of 
symbolic and algebraic computation, 
axiomatic logic system, mathematical pattern recognition, 
quantum chaos, and parallel computation.} 

\profile{Taksu Cheon}{%
received the B.S., M.S. and Ph.D. degrees in Physics 
from the University of Tokyo in 1980, 1982 and 1985, 
respectively. 
He is currently an Associate Professor in the 
Laboratory of Physics at Kochi University of Technology.  
His research interests are quantum mechanics, chaos, and 
quantum chaos.}

\end{document}